\documentclass[reqno]{amsart}
\usepackage{amssymb}
\usepackage{upref}

\newcommand{\bbN}{{\mathbb{N}}}

\newcommand{\bbZ}{{\mathbb{Z}}}
\newcommand{\bbC}{{\mathbb{C}}}

\newcommand{\calD}{{\mathcal D}}

\newcommand{\calK}{{\mathcal K}}

\newcommand{\no}{\nonumber}
\newcommand{\lb}{\label}
\newcommand{\bi}{\bibitem}
\newcommand{\f}{\frac}
\newcommand{\dott}{\,\cdot\,}
\newcommand{\hatt}{\widehat}  % use only for capital letters
\newcommand{\ti}{\widetilde} % use only for capital letters
\newcommand{\Oh}{O}

\DeclareMathOperator{\AKNS}{AKNS}

\DeclareMathOperator{\cBsq}{cBsq}
\newcommand{\Div}{\operatorname{Div}}

\newcommand{\uz}{{\underline{z}}}

\newcommand{\uxi}{{\underline{\Xi}}}
\newcommand{\n}{n}
\newcommand{\nn}{r}
\newcommand{\ul}{\underline}
\newcommand{\U}{\overline U}
\newcommand{\V}{\overline V}
\newcommand{\F}{\overline F}
\newcommand{\G}{\overline G}
\newcommand{\HH}{\overline H}
\newcommand{\ff}{\overline f}
\newcommand{\g}{\overline g}
\newcommand{\h}{\overline h}
\newcommand{\PPsi}{\overline \Psi}
\newcommand{\ppsi}{\overline \psi}
\newcommand{\pphi}{\overline \phi}
\newcommand{\mmu}{\overline \mu}
\newcommand{\nnu}{\overline \nu}
\allowdisplaybreaks
\numberwithin{equation}{section}

\newtheorem{theorem}{Theorem}[section]
\theoremstyle{definition}

\theoremstyle{remark}
\newtheorem{remark}[theorem]{Remark}

\begin{document}
\title[Classical Boussinesq hierarchy]{The classical 
Boussinesq hierarchy revisited} 

% Information for first author
\author[Gesztesy]{Fritz Gesztesy}
\address{Department of Mathematics,
University of Missouri,
Columbia, MO 65211, USA}
\email{fritz@math.missouri.edu}
\urladdr{http://www.math.missouri.edu/people/fgesztesy.html}
% Information for second author
\author[Holden]{Helge Holden}
\address{Department of Mathematical Sciences, Norwegian 
University of 
Science and Technology, N--7034 Trondheim, Norway}
\email{holden@math.ntnu.no}
\urladdr{http://www.math.ntnu.no/\~{}holden/}

\dedicatory{}
\thanks{Supported in part by the Research Council of Norway 
under grant
107510/410  and the
University of Missouri Research Board grant RB-97-086.}
\date{\today}

\begin{abstract}
We develop a systematic approach to the classical 
Boussinesq (cBsq)
hierarchy based on an elementary polynomial recursion 
formalism.
Moreover, the gauge equivalence between the cBsq and AKNS
hierarchies is studied in detail and used to provide an 
effortless derivation of algebro-geometric solutions 
and their theta function representations of the cBsq 
hierarchy. 
\end{abstract}

\maketitle

%%%%%%%%%%%%%%%%%%%%%%%%%%%%%%%%%%%%%%%%%%%%%%%%%%%%%%%%%%%
%----------- Section ----------------
\section{Introduction} \lb{intro}
%%%%%%%%%%%%%%%%%%%%%%%%%%%%%%%%%%%%%%%%%%%%%%%%%%%%%%%%%%

We develop an elementary algebraic approach to the classical
Boussinesq (cBsq) hierarchy in close analogy to previous
treatments of the KdV, AKNS, and Toda hierarchies
(cf.~\cite{BGHT98}, \cite{GH98}, \cite{GesztesyHolden:2000}, 
\cite{GRT96} and the references  therein). The
complete integrability of the classical Boussinesq system 
(and its
closely related variants, also known as the Kaup-Boussinesq,
Broer-Kaup, and classical Boussinesq-Burgers system), 
\begin{equation}
u_t+\f14 v_{xxx}+(uv)_x=0, \quad v_t+vv_x-u_x=0, \lb{1.1}
\end{equation}
was originally established by Kaup
\cite{Kaup:1975}, \cite{Kaup1:1975}. Various 
aspects of
this system (and its variants) are studied, for 
instance, in 
\cite{ClarksonLudlow:1994}, 
\cite{ConteMusettePickering:1994},
\cite{ConteMusettePickering:1995}, \cite{GengWu:1998},
\cite{Hirota:1985}, \cite{Hirota:1986}, \cite{Ito:1984}, 
\cite{JM77}, \cite{Kawamoto:1984},
\cite{Krishnan:1982}, \cite{Lax:1996},
\cite{LiuHuLi:1990}, \cite{MatveevYavor:1979}, 
\cite{Sachs:1988},
\cite{Smirnov:1986}, and \cite{Whitham:1974}.

Subsequently, the equivalence of the classical Boussinesq 
system
\eqref{1.1} and the AKNS system,
\begin{equation}
p_t+\f{i}{2}p_{xx}-ip^2q=0, \quad q_t-\f{i}{2}q_{xx}
+ipq^2=0, 
\lb{1.2}
\end{equation} 
was established by Jaulent and Miodek \cite{JM77} by 
means of the
explicit transformation 
\begin{equation}
u+\f{i}{2}v_x=-pq, \quad v=i\f{p_x}{p} \lb{1.3}
\end{equation}
(cf.~Section \ref{sect-equiv} for more details). 

The principal purpose of this note is threefold. First, we 
develop
the zero-curvature formalism of the cBsq hierarchy in
Section~\ref{sect-cBsq} using a polynomial recursion
formalism (independently of its connection with the AKNS
hierarchy). Second, we provide a new and elementary 
proof of the
gauge equivalence between the cBsq and AKNS hierarchies 
in
Section~\ref{sect-equiv}. Finally, using this gauge 
equivalence,
we derive the class of algebro-geometric solutions of 
the cBsq
hierarchy in Section~\ref{sect-algeb}.

The AKNS hierarchy and its class of algebro-geometric 
solutions,
the fundamental ingredients for Sections~\ref{sect-equiv} 
and
\ref{sect-algeb}, are briefly reviewed in 
Section~\ref{sect-AKNS}.

%%%%%%%%%%%%%%%%%%%%%%%%%%%%%%%%%%%%%%%%%%%%%%%%%%%%%%%%
%----------- Section ----------------
\section{The AKNS hierarchy} \lb{sect-AKNS}
%%%%%%%%%%%%%%%%%%%%%%%%%%%%%%%%%%%%%%%%%%%%%%%%%%%%%%%%%

In this section we review the construction of the AKNS 
hierarchy
and its algebro-geometric solutions following a recursive
approach to the AKNS zero-curvature formalism developed in
\cite{GesztesyRatnaseelan:1998}.

We start by recalling the recursive construction of the AKNS
hierarchy. Suppose
$p,q\colon
\bbC\to\bbC_\infty$ with
 $\bbC_\infty=\bbC\cup\{\infty\}$, are meromorphic and
 introduce the matrix
\begin{equation}
U(z,x) =  \begin{pmatrix}-iz & q(x) \\ p(x) &  iz 
\end{pmatrix}.
\lb{umatrix}
\end{equation}
Define
$\left\{f_{\ell}(x)\right\}_{\ell\in\bbN_0}$,
$\left\{g_{\ell}(x)\right\}_{\ell\in\bbN_0}$, and
$\left\{h_{\ell}(x)\right\}_{\ell\in\bbN_0}$
recursively by
\begin{align}
 f_0(x)&=-iq(x), \quad g_0(x)=1, \quad h_0(x)=ip(x),  
\no \\
 f_{\ell+1}(x)&= \f{i}{2} f_{\ell,x}(x) - 
i q(x) g_{\ell+1}(x),
\no \\
 g_{\ell+1,x}(x)&= p(x)f_{\ell}(x)+ q(x) h_{\ell}(x), 
\label{22} \\
 h_{\ell+1}(x)&= -\f{i}{2} h_{\ell,x}(x) + 
i p(x) g_{\ell+1}(x),
 \quad \ell\in\bbN_0. \no
\end{align}
Exlicitly, one finds
\begin{align}
f_0& = -iq,\quad f_1=\f12 q_x+c_1(-iq),\no \\
f_2& = \f{i}{4} q_{xx}-
      \f{i}{2}pq^2+ c_1\left( \f12 q_x\right)+c_2(-iq), 
\no \\
g_0& = 1,\quad g_1=c_1,\quad g_2=\f12 pq+c_2,\no \\
g_3& =  -\f{i}{4}(p_{_x}q - pq_x) + c_1\left(\f12 pq\right) 
+ c_3 ,
\label{3.3}\\ 
h_0& =i p,\quad h_1=\f12 p_{_x}+ c_1(ip), \no \\
h_2&=-\f{i}{4} p_{_{xx}}+\f{i}{2}p{^2}q+ c_1\left( \f12
p_{_x}\right)+c_2(i p), \no \\
&\text{ etc.}, \no
\end{align}
where $\{c_j\}_{j\in\bbN}\subset\bbC$ are integration 
constants.

Next, define the  matrix $V_{\n+1}(z,x)$ by
\begin{equation}
V_{\n+1}(z,x)=i \begin{pmatrix}
                  -G_{\n+1}(z,x) & F_{\n}(z,x) \\
-H_{\n}(z,x) & G_{\n+1}(z,x) \end{pmatrix},
\quad \n\in \bbN_0, \label{24}
\end{equation}
where $F_\n(z,x)$, $G_{\n+1}(z,x)$, and $H_\n(z,x)$ are
polynomials in $z\in \bbC$ of the type,
\begin{align}
F_\n(z,x)&=
\sum_{\ell=0}^{\n}f_{\n-\ell}(x)z^{\ell}
=-iq\prod_{j=1}^{\n}
(z-\mu_j(x)),
\no \\ G_{\n+1}(z,x)&=
\sum_{\ell=0}^{\n+1}g_{\n+1-\ell}(x)z^\ell,
\label{29}\\
H_\n(z,x)&=
\sum_{\ell=0}^{\n}h_{\n-\ell}(x)z^\ell=
ip\prod_{j=1}^{\n}
(z-\nu_j(x)). \no
\end{align}
Using the recursion \eqref{22} one verifies
\begin{align}
F_{n,x}&=-2izF_n+2qG_{n+1}, \no \\
G_{n+1,x}&=pF_n+qH_n, \lb{FGH} \\
H_{n,x}&=2izH_n+2pG_{n+1}, \no
\end{align}
implying
\begin{equation}
(G_{\n+1}^2 - F_\n H_\n)_x =0,
\label{213}
\end{equation}
and hence
\begin{equation}
G_{\n+1}(z,x)^2- F_\n(z,x) H_\n(z,x) = R_{2\n+2}(z),
\label{214}
\end{equation}
where $R_{2n+2}(z)$ is a monic polynomial of degree $2n+2$
with zeros
$\{E_0,\dots,E_{2n+1}\}\subset\bbC$. Thus,
\begin{equation}
R_{2n+2}(z) = \prod_{m=0}^{2n+1} (z-E_m),
\quad \{E_m\}_{m=0,\dots,2n +1}\subset \bbC.
\end{equation}
In particular, there is a naturally associated hyperelliptic 
curve
$\calK_n$ of genus $n$ obtained from the
characteristic equation for $V_{n+1}$,
\begin{align}
\det\left(yI-iV_{n+1}(z,x) \right)&=y^2-G_{n+1}(z,x)^2
+F_n(z,x)
H_n(z,x) \no \\
&= y^2-R_{2n+2}(z)=0,
\end{align}
with $I=\left(\begin{smallmatrix}1& 0 \\ 0 & 1
\end{smallmatrix}\right).$ We compactify the curve by 
adding two
points
$P_{\infty_\pm}$.  The compactified curve is still denoted
by $\calK_n$. A point $P$ on the curve
$\calK_n\backslash\{P_{\infty_\pm}\}$ is written as
$P=(z,y)$, where 
$y^2=R_{2n+2}(z)$. For  precise
definitions of detailed properties of the curve 
$\calK_n$ we 
refer to
Appendix A in \cite{GesztesyRatnaseelan:1998}.

The stationary AKNS hierarchy is obtained by enforcing the 
 stationary zero-curvature relation 
\begin{equation}
-V_{n+1,x}+[U,V_{\n+1}]=0,
\end{equation}
which, using \eqref{22}, \eqref{24}, and \eqref{29}, 
reduces to
\begin{align}
&-V_{n+1,x}+[U,V_{\n+1}] \no \\
&\quad =\begin{pmatrix}iG_{n+1,x}-ipF_n-iqH_n& -iF_{n,x}
+2zF_n+2iqG_{n+1}
\\ iH_{n,x}+2zH_n-2ipG_{n+1} & -iG_{n+1,x}+ipF_n
+iqH_n\end{pmatrix}
\lb{27a}\\
&\quad =\begin{pmatrix}ig_{n+1,x}-ipf_n-iqh_n 
& -2f_{n+1} \\ 
-2h_{n+1} & -ig_{n+1,x}+ipf_n+iqh_n \end{pmatrix}=0. 
\lb{27b}
\end{align}
Hence the stationary AKNS hierarchy is defined by
\begin{equation}
 \begin{pmatrix}h_{n+1}\\f_{n+1}\end{pmatrix}=0, 
\quad n\in\bbN_0.
\end{equation}

Explicitly, the first few equations read
\begin{align}
n=0:\quad  &\begin{pmatrix}
    - p_{_x} + c_1(-2 i p)  \\
   - q_x + c_1(2 i q) 
  \end{pmatrix}= 0, \no \\
n=1: \quad  &\begin{pmatrix} 
    \f{i}{2} p_{_{xx}}-i p{^2}q +
      c_1( - p_{_x})+c_2(-2 i p) \\[2mm]
    -\f{i}{2} q_{xx}+i pq^2+ c_1(-q_x)+c_2(2 i q)
  \end{pmatrix} = 0, \label{28} \\
& \text{ etc}. \no
\end{align}

Next we introduce a
deformation parameter $t_n\in\bbC$ in the functions $p$ 
and $q$, 
that is,
$p=p(x,t_n)$, $q=q(x,t_n)$.  
The time
dependent zero-curvature relation then reads
\begin{equation}
U_{t_n}-V_{n+1,x}+[U,V_{n+1}]=0.
\end{equation}
Employing \eqref{umatrix}, \eqref{27a}, and \eqref{27b}, one 
finds
\begin{align}
&U_{t_n}-V_{n+1,x}+[U,V_{n+1}] \no \\
&\quad=\begin{pmatrix}iG_{n+1,x}-ipF_n-iqH_n&
q_{t_n}-iF_{n,x}+2zF_n+2iqG_{n+1}
\\ p_{t_n}+iH_{n,x}+2zH_n-2ipG_{n+1} &
-iG_{n+1,x}+ipF_n+iqH_n\end{pmatrix}
\lb{28a}\\
&\quad=\begin{pmatrix} ig_{n+1,x}-ipf_n-iqh_n
& q_{t_n}-2f_{n+1}\\
p_{t_n}-2 h_{n+1}&-ig_{n+1,x}+ipf_n+iqh_n 
 \end{pmatrix}=0. \lb{28b}
\end{align}
Hence the AKNS hierarchy is defined by
\begin{equation}
\AKNS_n(p,q)=\begin{pmatrix}p_{t_n}-2h_{n+1} 
\\ q_{t_n}-2f_{n+1} \end{pmatrix}=0, \quad n\in\bbN_0. 
\lb{28c}
\end{equation}

Explicitly, the first few equations read
\begin{align}
 \AKNS_0(p,q)&=\begin{pmatrix} p_{t_{0}}- p_{x} +
 c_1(-2 i p)  \\ q_{t_{0}} - q_x + c_1(2 i q) 
\end{pmatrix}
= 0, \no   \\
\AKNS_1(p,q)&=\begin{pmatrix}
         p_{t_{1}}  +\f{i}{2} p_{_{xx}}-i p{^2}q +
    c_1\left( - p_{_x}\right)
    +c_2(-2 i p) \\[2mm]
      q_{t_{1}}-\f{i}{2} q_{xx}+i pq^2
    + c_1\left(- q_x\right)+c_2(2 i q)\end{pmatrix}=0,
\lb{28d} \\
&\text{  etc.} \no
\end{align}

We also recall a scale invariance of the AKNS 
hierarchy (see
\cite{GesztesyRatnaseelan:1998} for details). 
Suppose $(p,q)$ satisfies one of the AKNS equations 
\eqref{28c}
for some $n\in\bbN_0$,
\begin{equation}
\AKNS_n(p,q)=0
\label{28e}
\end{equation}
and consider the scale transformation
\begin{equation}
(p(x,t_n),q(x,t_n))\to ({\breve p}(x,t_n),
{\breve q}(x,t_n))=
(A p(x,t_n),A^{-1}q(x,t_n)),\; \; A\in \bbC\backslash
\{0\}.
\label{28f}
\end{equation}
Then
\begin{equation}
\AKNS_n({\breve p},{\breve q})=0.
\label{28g}
\end{equation}

The outlined recursive approach is not confined to the 
hierarchy
of AKNS evolution equations. Analogous considerations 
apply 
to the
KdV, Toda, Boussinesq hierarchy, etc. (see 
\cite{BGHT98}, 
\cite{GH98}, \cite{GesztesyHolden:2000}, 
\cite{GRT96} and the references  therein).

Next we turn to the class of algebro-geometric 
solutions 
of the
AKNS hierarchy.  For brevity we present the 
formulas in the 
time-dependent setting  only. The corresponding 
stationary
formulas easily follow as a special case
(cf.~\cite{GesztesyRatnaseelan:1998}). Let
$(p^{(0)},q^{(0)})$ be a solution of the
$n$th stationary AKNS system, that is,
\begin{equation}
f_{n+1}(p^{(0)},q^{(0)})=h_{n+1}(p^{(0)},q^{(0)})=0,
\end{equation}
for a given set $\{c_j\}_{j=1,\dots,n+1}\subset\bbC$ of 
integration
constants. Consider subsequently the $r$th time-dependent 
AKNS system for some fixed
$r\in\bbN_{0}$, with integration constants  $\{\tilde
c_j\}_{j=1,\dots,r+1}\subset\bbC$. The
corresponding quantities $f_j$, $g_j$, etc.\ (that 
is, with 
$c_1,\dots,c_{r+1}$
replaced by $\tilde c_1,\dots,\tilde c_{r+1}$), for 
this system 
will be
denoted with a tilde, $\tilde f_j$, $\tilde g_j$, 
$\tilde
h_j,$ $\widetilde V_{r+1},$ etc. Thus, we are 
interested in the 
construction of solutions  $(p,q)$ of 
\begin{equation}
\AKNS_r(p,q)=\begin{pmatrix}p_{t_r}-2\tilde h_{r+1} \\ 
q_{t_r}-2\tilde f_{r+1} \end{pmatrix}=0,\quad 
(p,q)|_{t_r=t_{0,r}}=(p^{(0)},q^{(0)}).
\end{equation}  
These  algebro-geometric AKNS solutions are obtained by a 
careful analysis of
two specific functions on $\calK_n$.
First one defines the meromorphic function $\phi$ on 
$\calK_n$ by
\begin{multline}
\phi(P,x,t_\nn)=\f{y(P)+G_{\n+1}(z,x,t_\nn)}
{F_\n(z,x,t_\nn)}=
     \f{-H_\n(z,x,t_\nn)}{y(P)-G_{\n+1}(z,x,t_\nn)},  \\
 P=(z,y) \in \calK_\n. \label{415}
\end{multline}
The divisor of $\phi(P,x,t_n)$ is given by
\begin{equation}
(\phi(P,x,t_\nn)) = \calD_{P_{\infty_+}
{\ul {\hat \nu}}(x,t_\nn)} -
\calD_{P_{\infty_-}{\ul
{\hat \mu}}(x,t_\nn)}.
\end{equation}
Here
\begin{multline}
{\ul {\hat \mu}}(x,t_\nn) =
(\hat \mu_1(x,t_\nn), \dots ,\hat \mu_n(x,t_\nn))
\in \sigma^n \calK_n, \\
\hat \mu_j(x,t_\nn) =
(\mu_j(x,t_\nn), G_{n+1}(\mu_j(x,t_\nn),x,t_\nn)), \quad
j=1,\dots, n,
\end{multline}
and
\begin{multline}
{\ul {\hat \nu}}(x,t_\nn)
=(\hat \nu_1(x,t_\nn), \dots ,\hat \nu_n(x,t_\nn))
\in \sigma^n \calK_n,\\
\hat \nu_j(x,t_\nn) = (\nu_j(x,t_\nn),
-G_{n+1}(\nu_j(x,t_\nn),x,t_\nn)),
\quad j=1,\dots, n,
\end{multline}
where $\sigma^n\calK_n$ denotes the $n$th symmetric 
power of
$\calK_n$ and  
\begin{multline} \lb{A.17}
\calD_{Q_0\ul Q}(P)=\calD_{Q_0}+\calD_{\ul Q}, 
\quad \calD_{\ul
Q}=\calD_{Q_1}+\cdots +\calD_{Q_n}, \\
  {\ul Q}=(Q_1, \dots ,Q_n) \in \sigma^n \calK_n,
\quad Q_0\in\calK_n,
\end{multline}
and for any $Q\in\calK_n$,
\begin{equation} \lb{A.18}
\calD_Q \colon  \calK_n \to\bbN_0, \quad
P \mapsto  \calD_Q (P)=
\begin{cases} 1 & \text{for $P=Q$},\\
0 & \text{for $P\in \calK_n\backslash \{Q\}$}. \end{cases}
\end{equation}
Secondly, we introduce
the Baker--Akhiezer vector,
\begin{align}
\Psi(P,x,x_0,t_\nn,t_{0,\nn})&=\begin{pmatrix}
             \psi_1(P,x,x_0,t_\nn,t_{0,\nn}) \\
              \psi_2(P,x,x_0,t_\nn,t_{0,\nn})
     \end{pmatrix}, \label{421} \\
\psi_1(P,x,x_0, t_\nn,t_{0,\nn}) &=
\exp \left( \int_{x_0}^x dx'\, (-iz+
q(x',t_\nn)\phi(P,x',t_\nn))  \right. \no \\
&\qquad
 \left. +i\int_{t_{0,\nn}}^{t_\nn} \, ds (\ti F_\n (z,x_0,
s) \phi (P,x_0,s) -\ti G_{\n+1} (z,x_0, s)) \right), \no \\
\psi_2(P,x,x_0,t_\nn,t_{0,\nn})&=
\phi(P,x,t_\nn)\psi_1(P,x,x_0,t_\nn,t_{0,\nn}), \no\\
& \quad \quad \quad \quad \quad \quad \quad \quad P\in
\calK_\n\backslash
\{P_{\infty_{\pm}}\}, \,
(x,t_\nn)\in\bbC^2.\no
\end{align}
 
The functions $\phi$ and $\psi$ satisfy the  following 
fundamental
properties.

%-----------theorem 
\begin{theorem}[see \cite{GesztesyHolden:2000}, 
\cite{GesztesyRatnaseelan:1998}]
\label{l41} 
Let $P=(z,y) \in \calK_n 
\backslash \{P_{\infty_{\pm}}\}$ and  
$(z,x,x_0,t_r,t_{0,r}) \in \bbC^5$. Then
$\phi(P,x,t_r)$   satisfies  
\begin{align}
&\phi_{x}(P,x,t_r)+q(x,t_r)\phi(P,x,t_r)^2- 
2iz\phi(P,x,t_r)
     =p(x,t_r). \label{424}
\end{align}
$\Psi(\dott,x,x_0,t_0,t_{0,r})$ is meromorphic on 
$\calK_n\backslash\{P_{\infty_\pm} \}$ and 
\begin{align}
\Psi_x(P,x,x_0, t_r,t_{0,r})&=
U(z,x,t_r)\Psi(P,x,x_0,t_r,t_{0,r}),  \label{431}\\
iy(P)\Psi(P,x,x_0, t_r,t_{0,r})
&=V_{n+1}(z,x,t_r)\Psi(P,x,x_0, t_r,t_{0,r}), 
\lb{431aa} \\
\Psi_{t_r}(P,x,x_0, t_r,t_{0,r})
&=\ti V_{r+1}(z,x,t_r)\Psi(P,x,x_0, t_r,t_{0,r}). 
\lb{431aaa}
\end{align}
\end{theorem}
%---------- end of theorem

Further properties of $\phi$ and $\Psi$ can be found in 
\cite{GesztesyRatnaseelan:1998}, Lemma 4.1.

In order to express the basic quantities in terms of 
Riemann's theta function associated with $\calK_n$, we 
need to
introduce differentials and some more notation in connection
with the hyperelliptic curve $\calK_n$. In the following 
we assume  $\calK_n$ to be nonsingular, that is,
\begin{equation}
E_m \neq E_{m'}, \quad m, m'=0,\dots,2 n+1,\, m\neq m'.
\lb{333}
\end{equation}
Given a canonical homology basis 
$\{a_j,b_j\}_{j=1,\dots,n}$ for $\calK_n$ with 
intersection matrix $a_j\circ b_k=\delta_{j,k},$ one 
denotes by $\omega_j$, $j=1,\dots,n$ a normalized basis 
of the space of holomorphic differentials on $\calK_n,$ 
\begin{equation}
\int_{a_k} \omega_j=\delta_{j,k}, \quad j,k=1,\dots,n.
\end{equation}
In addition, one considers a canonical dissection of 
$\calK_n$ along its cycles yielding the simply connected 
interior $\widehat \calK_n$ of the fundamental polygon
\begin{equation} 
\partial\widehat\calK_n=
a_1b_1a_1^{-1}b_1^{-1}\dots a_n^{-1}b_n^{-1}. 
\end{equation}
Next, 
we choose, without loss of generality, the base point
$P_0=(E_0,0)\in\calK_n$ and denote by
$\ul A_{P_0}$, $\ul{\alpha}_{P_0}$ the Abel maps
\begin{align}
\ul A_{P_0} \colon 
\calK_n \to J(\calK_n),\quad
P\mapsto \ul A_{P_0} (P) &=(A_{P_0,1} (P),
\dots,A_{P_0,n} (P)) 
\no \\
&=\left(\int_{P_{0}}^P \omega_1,\dots,\int_{P_{0}}^P 
\omega_n\right) \text{(mod $L_n$)}, \lb{aa46}
\end{align}
and
\begin{equation}
\ul \alpha_{P_0}  \colon
\Div(\calK_n) \to J(\calK_n),\quad
\calD \mapsto \ul \alpha_{P_0} (\calD)
=\sum_{P \in \calK_n} \calD (P) \ul A_{P_0} (P),
\lb{aa47a}
\end{equation}
where  
\begin{equation}
L_n=\{\uz\in\bbC^n \mid \uz=\underline{N}+
\tau\underline{M}, \;
\underline{N},\underline{M}\in\bbZ^n\} \lb{aa45}
\end{equation}
denotes the period lattice, and
\begin{equation}
J(\calK_n)=\bbC^n/L_n  \lb{aa45a}
\end{equation}
the Jacobi variety. In addition we introduce 
\begin{align}
&\ul {\widehat A}_{P_0} \colon 
\widehat \calK_n \to \bbC^n,\quad
P\mapsto \ul {\widehat A}_{P_0} (P) =
(\widehat A_{P_0,1} (P),
\dots,\widehat A_{P_0,n} (P))  \no \\
&\hspace*{4.77cm} =\left(\int_{P_{0}}^P 
\omega_1,\dots,\int_{P_{0}}^P 
\omega_n\right), \lb{aa47} \\
& \hat{\ul \alpha}_{P_0}  \colon
\Div(\calK_n) \to\bbC^n,\quad
\calD \mapsto \hat{\ul \alpha}_{P_0} (\calD)
=\sum_{P \in \calK_n} \calD (P) 
\ul {\widehat A}_{P_0} (P), \lb{aa47b}
\end{align}
and  $\ul \Xi_{P_0}=(\Xi_{P_0,1},\dots,\Xi_{P_0,n})$, 
the vector of Riemann constants, by
\begin{align}
&{\ul \Xi}_{P_0}= 
\widehat {\ul \Xi}_{P_0} \text{(mod $L_n$)}, \no \\
& \widehat \Xi_{P_{0},j}=
\frac12\left(1+\int_{b_j}\omega_j\right)-
\sum_{\substack{\ell=1 \\ \ell\neq j}}^n\int_{a_\ell}
\omega_\ell(P)\widehat A_{P_0,j}(P), 
 \quad j=1,\dots,n. \lb{aa55}
\end{align}
Next, consider the normal differential of the third
kind
$\omega^{(3)}_{P_{\infty_+}, P_{\infty_-}}$ with 
simple poles
at $P_{\infty_+}$ and
$P_{\infty_-}$, corresponding residues $+1$ and $-1$,
vanishing $a$-periods,
being holomorphic otherwise on $\calK_n$.  Hence one 
obtains
\begin{align}
& \omega^{(3)}_{P_{\infty_+}, P_{\infty_-}} =
\f{\prod_{j=1}^n (\tilde\pi -\lambda_j)\,
d\tilde\pi}{y},
\quad \omega^{(3)}_{P_{\infty_-}, P_{\infty_+}}
=-\omega^{(3)}_{P_{\infty_+}, P_{\infty_-}},
\lb{335}\\
& \int_{a_j} \omega^{(3)}_{P_{\infty_+}, 
P_{\infty_-}} = 0,
\quad j=1,\dots, n,
\lb{336}
\end{align}
\begin{align}
&\ul U^{(3)} =(U_1^{(3)},\dots ,U_n^{(3)}),\no \\
&U_j^{(3)} = \f{1}{2\pi i}
\int_{b_j} \omega_{P_{\infty_+}, P_{\infty_-}}^{(3)}
=\hatt A_{P_{\infty_-}, j} (P_{\infty_+})
   = 2 \hatt A_{P_0, j}(P_{\infty_+}), \quad j=1,\dots,n, 
\lb{337}
\end{align}
\begin{align}
\int_{P_0}^P \omega_{P_{\infty_+}, P_{\infty_-}}^{(3)}
\underset{\zeta \to 0}{=}\pm (\ln(\zeta)-\ln(\omega_0)
+\Oh(\zeta)), \text{  $P=(\zeta^{-1},y)$ near 
$P_{\infty_{\pm}}$},
\lb{338}
\end{align}
where the numbers $\{ \lambda_j \}_{j=1,\dots,n}$
are determined by the
normalization \eqref{336}. The Abelian differentials 
of the
second kind $\omega_{P_{\infty_{\pm}},0}^{(2)}$ are 
chosen 
such that
\begin{equation}
\omega_{\infty_\pm,0}^{(2)}
\underset{\zeta \to 0}{=}(\zeta^{-2} +\Oh(1))
\, d\zeta \text{ near $P_{\infty_\pm}$},
\lb{339}
\end{equation}
\begin{equation}
 \int_{a_j} \omega_{P_{\infty_{\pm}},0}^{(2)} = 0,
\quad j=1,\dots,n,
\lb{340}
\end{equation}
\begin{align}
&\ul U_0^{(2)} =(U_{0,1}^{(2)},\ldots, U_{0,n}^{(2)}),\; \;
U_{0,j}^{(2)} = \f{1}{2\pi i}
\int_{b_j} \Omega_{0}^{(2)},\quad
\Omega_{0}^{(2)}=\omega_{P_{\infty_{+}},0}^{(2)}
-\omega_{P_{\infty_{-}},0}^{(2)},
\lb{341}
\end{align}
\begin{align}
\int_{P_0}^P \Omega_{0}^{(2)}
\underset{\zeta \to 0}{=}\mp (\zeta^{-1}+e_{0,0}
+e_{0,1}\zeta + \Oh(\zeta^2)), \; \; P=(\zeta^{-1},y)\; \;
\text{near}\; \; P_{\infty_{\pm}}.
\lb{342}
\end{align}
In addition we define
Abelian differentials
of the second kind $\omega_{P_{\infty_{\pm}},r}^{(2)}$ by
\begin{align}
\omega_{P_{\infty_\pm},r}^{(2)}
&\underset{\zeta \to 0}{=}(\zeta^{-2-r} +\Oh(1))
\, d\zeta \text{ near $P_{\infty_\pm}$, $r\in\bbN_0$},
\lb{452} \\
 \int_{a_j} \omega_{P_{\infty_{\pm}},r}^{(2)} &= 0,
\quad j=1,\dots,n, \lb{453} \\
\ti{\ul U}_r^{(2)} &=(\ti U_{r,1}^{(2)},
\dots,\ti U_{r,n}^{(2)}),\quad
\ti U_{r,j}^{(2)} = \f{1}{2\pi i}
\int_{b_j}\ti \Omega_{r}^{(2)},\no \\
\ti \Omega_{r}^{(2)}&= \sum_{q=0}^r (q+1)\tilde c_{r-q}
(\omega_{P_{\infty_+},q}^{(2)}-
\omega_{P_{\infty_-},q}^{(2)}),
\lb{454} \\
\int_{P_0}^P\ti \Omega_{r}^{(2)}
&\underset{\zeta \to 0}{=}\mp \left(\sum_{q=0}^r 
\tilde c_{r-q}
\zeta^{-1-q}+\tilde e_{r,0}
+\Oh(\zeta)\right), \quad P=(\zeta^{-1},y) 
\text{  near $P_{\infty_{\pm}}$}, \lb{455}
\end{align}
with $\{\tilde c_\ell\}_{\ell=1,\dots,r}\subset\bbC$, 
$\tilde
c_0=1$ denoting integration constants.

Finally, we abbreviate
\begin{align}
 \uz(P,\ul Q)&= \ul {\widehat A}_{P_0}(P)-\hat{\ul 
\alpha}_{P_0}(\calD_{\ul Q})-
\widehat {\uxi}_{P_0}, \lb{345} \\
\uz_{\pm}(\ul Q)&=\uz(P_{\infty_{\pm}},\ul Q),\quad
\ul Q=(Q_1,\dots,Q_n).
\lb{346}
\end{align}
Next, assume that $(p,q)$ satisfies the $r$th 
time-dependent 
AKNS
equation with initial data 
$(p^{(0)}, q^{(0)})$ being solutions of the $n$th 
stationary AKNS
equation at $t=t_{0,r}$, that is,
\begin{equation}
(p(x,t_{0,r}),q(x,t_{0,r}))=(p^{(0)}(x),q^{(0)}(x)), 
\quad x\in\bbC.
\end{equation}
The principal aim is to derive explicit formulas for 
the solution $(p,q)$ as well as
the function $\phi$ and the  Baker--Akhiezer 
function $\Psi$ 
in terms of the Riemann theta function
\begin{align}
\theta (\ul z) &=\sum_{\ul m \in\bbZ^n}
 \exp \big(2\pi i (\ul m,\ul z) + \pi i (\ul m,
\tau \ul m)\big), \quad \ul z \in\bbC^n, \no \\
(\ul u, \ul v)&=\sum_{j=1}^n \overline{u}_j v_j,  \quad
\tau=\left(\int_{b_j} \omega_k\right)_{j,k=1,\dots,n}.
\lb{aa49}
\end{align}
This is the content of the next theorem.
%-------------- theorem ----------------
\begin{theorem}[see
\cite{GesztesyHolden:2000}, 
\cite{GesztesyRatnaseelan:1998}] \lb{t2.2}
Let $P\in\calK_n\backslash\{P_{\infty_\pm}\}$, 
$(x,x_0,t_r,t_{0,r})\in\bbC^4$,
assume $\calK_n$ to be nonsingular, and suppose 
$\calD_{\ul {\hat \mu}(x,t)}$, or 
equivalently,
$\calD_{\ul {\hat \nu}(x,t)}$ to be nonspecial, that is,  
their
index of speciality vanishes, $i(\calD_{\ul 
{\hat \mu}(x,t)})=
i(\calD_{\ul {\hat \nu}(x,t)})=0$. 
Moreover, suppose that $(p,q)$ satisfies $\AKNS_r(p,q)=0$  
with $(p,q)|_{t_r=t_{0,r}}=(p^{(0)},q^{(0)})$, a solution
of the $n$th stationary AKNS system.
Let  $\phi$ and $\Psi$ be defined by
\eqref{415} and \eqref{421}, respectively.
Then
\begin{align}
&\phi(P,x,t_r)=\f{2i}{q(x_0,t_{0,r})\omega_0}
\f{\theta(\uz_-(\ul {\hat \mu}(x_0,t_{0,r})))}
{\theta(\uz_+(\ul {\hat \mu}(x_0,t_{0,r})))}
\f{\theta(\uz_+(\ul {\hat \mu}(x,t_r)))}
{\theta(\uz_-(\ul {\hat \nu}(x,t_r)))}
\f{\theta(\uz(P,\hat{\ul \nu}(x,t_{r})))}
{\theta(\uz(P,\ul {\hat \mu}(x,t_{r})))}\times \no \\
&\hspace*{1.9cm} \times
\exp\big(\int_{P_0}^P\omega_{P_{\infty_+}, P_{\infty_-}}^{(3)}
-2i(x-x_0)e_{0,0}-2i(t_r-t_{0,r})\tilde e_{r,0} \big),
\lb{465} \\
&\psi_1(P,x,x_0,t_r,t_{0,r})=
\f{\theta(\uz_+(\ul {\hat \mu}(x_0,t_{0,r})))}
{\theta(\uz(P,\ul {\hat \mu}(x_0,t_{0,r})))}
\f{\theta(\uz(P,\hat{\ul \mu}(x,t_r)))}
{\theta(\uz_+(\ul {\hat \mu}(x,t_r)))} \times \lb{466} \\
&\hspace*{2.4cm} \times
\exp\bigg(i(x-x_0)\big(e_{0,0}
+\int_{P_0}^P\Omega_0^{(2)}\big)
+i(t_r-t_{0,r})
\big(\tilde e_{r,0}+\int_{P_0}^P{\ti 
\Omega}_r^{(2)}\big)\bigg),
\no \\
&\psi_2(P,x,x_0,t_r,t_{0,r})=
\f{2i}{q(x_0,t_{0,r})\omega_0}
\f{\theta(\uz_-(\ul {\hat \mu}(x_0,t_{0,r})))}
{\theta(\uz(P,\ul {\hat \mu}(x_0,t_{0,r})))}
\f{\theta(\uz(P,\ul {\hat \nu}(x,t_r)))}
{\theta(\uz_-(\ul {\hat \nu}(x,t_r)))}\times \no \\
&\hspace*{3.3cm} \times
\exp\bigg(\int_{P_0}^P\omega_{P_{\infty_+}, 
P_{\infty_-}}^{(3)}+
i(x-x_0)\big(-e_{0,0}+
\int_{P_0}^P\Omega_0^{(2)}\big)\no \\
&\hspace*{3.3cm} +i(t_r-t_{0,r})
\big(-\tilde e_{r,0}+\int_{P_0}^P{\ti 
\Omega}_r^{(2)}\big)\bigg). 
\lb{467}
\end{align}
Furthermore, one derives
\begin{align}
p(x,t_r)&=p(x_0,t_{0,r})
\f{\theta(\uz_-(\ul {\hat \nu}(x_0,t_{0,r})))}
{\theta(\uz_+(\ul {\hat \nu}(x_0,t_{0,r})))}
\f{\theta(\uz_+(\ul {\hat \nu}(x,t_r)))}
{\theta(\uz_-(\ul {\hat \nu}(x,t_r)))} \times \no \\
&\quad \times
\exp(-2i(x-x_0)e_{0,0}-2i(t_r-t_{0,r})\tilde e_{r,0}), 
\lb{468} \\
q(x,t_r)&=q(x_0,t_{0,r})
\f{\theta(\uz_+(\ul {\hat \mu}(x_0,t_{0,r})))}
{\theta(\uz_-(\ul {\hat \mu}(x_0,t_{0,r})))}
\f{\theta(\uz_-(\ul {\hat \mu}(x,t_r)))}
{\theta(\uz_+(\ul {\hat \mu}(x,t_r)))}\times \no \\
& \quad\times
\exp(2i(x-x_0)e_{0,0}+2i(t_r-t_{0,r})\tilde e_{r,0}),
\lb{469} \\
p(x_0,t_{0,r})q(x_0,t_{0,r})&=\f{4}{\omega_0^2}
\f{\theta(\uz_+(\ul {\hat \nu}(x_0,t_{0,r})))}
{\theta(\uz_-(\ul {\hat \nu}(x_0,t_{0,r})))}
\f{\theta(\uz_-(\ul {\hat \mu}(x_0,t_{0,r})))}
{\theta(\uz_+(\ul {\hat \mu}(x_0,t_{0,r})))},
\lb{470}\\
p(x,t_r)q(x,t_r)&=-e_{0,1} -
\f{\partial^2}{\partial x^2}
\ln(\theta(\uz_+(\hat {\ul \mu}(x,t_r)))),
 \lb{483} \\
\intertext{and}
 {\ul \alpha}_{P_0}(\calD_{\ul {\hat \mu}(x,t_r)})&=
  {\ul \alpha}_{P_0}(\calD_{\ul {\hat \mu}(x_0,t_{0,r})})
-i(x-x_0)
   \ul U_0^{(2)}-i(t_r-t_{0,r})\ti{\ul U}_r^{(2)},
 \lb{471} \\
  {\ul \alpha}_{P_0}(\calD_{\ul {\hat \nu}(x,t_r)})&=
  {\ul \alpha}_{P_0}(\calD_{\ul {\hat \nu}(x_0,t_{0,r})})
-i(x-x_0)
   \ul U_0^{(2)}-i(t_r-t_{0,r})\ti{\ul U}_r^{(2)}.
 \lb{472}
\end{align}
\end{theorem}
%---------- end of theorem ---------

Algebro-geometric solutions of the AKNS equations (i.e., 
the case
$r=1,$ $n\in\bbN$) have previously been derived by a 
variety of
authors, see, for instance, \cite{BBEIM94}, \cite{Du77},
\cite{Du83}, \cite{It81}, \cite{It86}, \cite{Kr77}, 
\cite{MA81},
\cite{Me87}, \cite{Pr85} and the literature cited 
therein. The 
principal contribution of \cite{GesztesyRatnaseelan:1998} 
to this
circle of ideas is an effortless treatment of 
algebro-geometric
solutions of the entire AKNS hierarchy (i.e., for 
$r,n\in\bbN$)
using the elementary polynomial recursion formalism 
outlined in
the first part of this section.

%%%%%%%%%%%%%%%%%%%%%%%%%%%%%%%%%%%%%%%%%%%%%%%%%%%%%%%%%%%
%----------- section
\section{The classical Boussinesq hierarchy} \lb{sect-cBsq}
%%%%%%%%%%%%%%%%%%%%%%%%%%%%%%%%%%%%%%%%%%%%%%%%%%%%%%%%%%

In this section we follow the zero-curvature formalism 
introduced
by Geng and Wu \cite{GengWu:1998} for the cBsq hierarchy 
and adapt
it to the recursion formalism outlined in 
Section~\ref{sect-AKNS}. Fix\footnote{The constants
$\alpha$ and
$\beta$  remain fixed  in the following and will not be 
emphasized
in the notation.}
$\alpha\in\bbC\backslash\{0\},$ $\beta\in\bbC$, and define
the matrix
\begin{equation}
	\U(z,x)=\begin{pmatrix}-iz-\alpha v(x) & u(x)
+\beta v_{x}(x)\\
 -1 & iz+\alpha v(x) \end{pmatrix}. 
\lb{u-matrix}
\end{equation}
Define recursively
$\{\ff_{j}(x)\}_{j\in\bbN_{0}}$, 
$\{\g_{j}(x)\}_{j\in\bbN_{0}}$, 
and 
$\{\h_{j}(x)\}_{j\in\bbN_{0}}$ by
\begin{subequations} \lb{recurs}
\begin{align} 
\ff_{0}(x)&=-i(u(x)+\beta v_{x}(x)), \quad \g_{0}(x)=1, 
\quad \h_{0}(x)=-i, 
\lb{recurs-a}\\
\ff_{j+1}(x)&=\f{i}{2} \ff_{j,x}(x)+i \alpha v(x) 
\ff_{j}(x)
-i (u(x)+\beta v_{x}(x))\g_{j+1}(x),
\lb{recurs-b}\\
\g_{j+1,x}(x)&=(u(x)+\beta v_{x}(x)) \h_{j}(x)
-\ff_{j}(x),
\lb{recurs-c}\\
\h_{j+1}(x)&=-\f{i}{2} \h_{j,x}(x)+i \alpha v(x) 
\h_{j}(x)-i
\g_{j+1}(x), 
\quad j\in\bbN_0. \lb{recurs-d}
\end{align}
\end{subequations}
Explicitly, the first few elements read
\begin{align}
\ff_{1}&=\f12(u+\beta 
v_{x})_x + \alpha v (u+\beta v_{x})+c_1(-i)(u+\beta v_x), 
\no \\
\g_{1}&=c_{1}, \quad 
\g_{2}=-\f12(u+\beta v_{x})+c_{2}, \lb{ex} \\
\h_{1}&=\alpha v-i c_{1}, \no \\
&\text{ etc.} \no
\end{align}
where $\{c_{j}\}_{j\in\bbN}\subset\bbC$ are integration
constants. 

Next, define
\begin{equation}
\V_{n+1}(z,x) =i\begin{pmatrix} -\G_{n+1}(z,x) 
& \F_{n}(z,x) \\
-\HH_{n}(z,x) & \G_{n+1}(z,x) \end{pmatrix} \lb{v-matrix}
\end{equation}
where 
$\F_\n(z,x)$, $\G_{\n+1}(z,x)$, and $\HH_\n(z,x)$ are
polynomials in $z\in \bbC$,
\begin{align}
\F_\n(z,x)&=
\sum_{\ell=0}^{\n}\ff_{\n-\ell}(x)z^{\ell}
=-i(u(x)+\beta v_x(x))\prod_{j=1}^{\n}(z-\mmu_j(x)),
\no \\ \G_{\n+1}(z,x)&=
\sum_{\ell=0}^{\n+1}\g_{\n+1-\ell}(x)z^\ell,
\label{29aa}\\
\HH_\n(z,x)&=\sum_{\ell=0}^{\n}\h_{\n-\ell}(x)z^\ell=
-i\prod_{j=1}^{\n}(z-\nnu_j(x)). \no
\end{align}
Using the recursion \eqref{recurs} one verifies
\begin{align}
\F_{n,x}&=-2(iz+\alpha v)\F_{n}+2(u+\beta v_{x})\G_{n+1}, 
\no \\
\G_{n+1,x}&=(u+\beta v_{x})\HH_{n}-\F_{n}, \lb{bar-FGH}\\
\HH_{n,x}&=2(iz+\alpha v) \HH_{n}-2\G_{n+1}, \no
\end{align}
implying
\begin{equation}
(\G_{n+1}^2-\F_n\HH_n)_x=0
\end{equation}
and hence 
\begin{equation}
\G_{\n+1}(z,x)^2- \F_\n(z,x) \HH_\n(z,x) = R_{2\n+2}(z),
\label{214a}
\end{equation}
where $R_{2n+2}(z)$ is a monic polynomial of degree $2n+2$
with zeros $\{E_0,\dots,E_{2n+1}\}\subset\bbC$. Thus,
\begin{equation}
R_{2n+2}(z) = \prod_{m=0}^{2n+1} (z-E_m),
\quad \{E_m\}_{m=0,\dots,2n +1}
\subset \bbC.
\end{equation}
Again the corresponding hyperelliptic curve
$\calK_n$ of genus $n$ is naturally obtained from the
characteristic equation for $\V_{n+1}$,
\begin{align}
\det\left(yI-i\V_{n+1}(z,x) \right)&=y^2-\G_{n+1}(z,x)^2
+\F_n(z,x)
\HH_n(z,x) \no \\
&= y^2-R_{2n+2}(z)=0.
\end{align}
The corresponding zero-curvature relation then reads,
employing \eqref{v-matrix} and \eqref{29aa}, 
\begin{align}
&-\V_{n+1,x}+[\U,\V_{n+1}] \lb{30a}\\
&\quad=i\left(\begin{smallmatrix} \G_{n+1,x}
+\F_n-(u+\beta v_x)\HH_n &
-\F_{n,x}-2(iz+\alpha v)\F_n+2(u+\beta v_x)\G_{n+1}
\\ \HH_{n,x}-2(iz+\alpha v)\HH_n+2\G_{n+1} &
-\G_{n+1,x}-\F_n+(u+\beta v_x)\HH_n 
\end{smallmatrix}\right)=0.\no 
\end{align}
Using the recursion \eqref{recurs} to compute 
$\ff_j$, $\g_j$, and
$\h_j$ for $j=0,\dots,n$, \eqref{30a} equals 
\begin{align}
&-\V_{n+1,x}+[\U,\V_{n+1}] \lb{30b} \\
&\quad=i\begin{pmatrix} \g_{n+1,x}
+\ff_n-(u+\beta v_x)\h_n 
& -\ff_{n,x}-2\alpha v \ff_n+2(u+\beta v_x)\g_{n+1} \\ 
\h_{n,x}-2\alpha v \h_n+2\g_{n+1} & 
-\g_{n+1,x}-\ff_n+(u+\beta v_x)\h_n \end{pmatrix} 
=0. \no
\end{align}
Next, let
\begin{equation}
\g_{n+1}=-\f12 \h_{n,x}+\alpha v\h_n \lb{30aaa}
\end{equation}
(consistent with $\h_{n+1}=0$ in \eqref{recurs-d}).
Inserting \eqref{30aaa} into \eqref{30b}, we find that 
the stationary cBsq 
 hierarchy is given by 
\begin{equation}
 \begin{pmatrix}  \ff_{n,x}+2\alpha v \ff_n
+(u+\beta v_x)(\h_{n,x}-2\alpha v\h_n)\\
-\f12\h_{n,xx}+\alpha(v\h_n)_x +\ff_{n}-(u+\beta v_x)\h_n 
\end{pmatrix}=0, 
\quad n\in\bbN_0.
\end{equation}

\begin{remark}
Observe that due to \eqref{30aaa}, the $n$th 
stationary cBsq system
will contain only integration constants $c_1,\dots,c_n$ 
for $n\in\bbN$
coming from integrating \eqref{recurs-c}. Since we have
$\g_{n+1,x}+\ff_n-(u+\beta v_x)\h_n=0$ from
\eqref{30b}, our definition \eqref{30aaa} is consistent 
with the
definition of
$\g_{n+1}$ given by the recursion \eqref{recurs-c}.  
However, no new
integration constant is introduced. 
\end{remark}

The first few equations (after some simplifications) read
\begin{align}
n=0: \quad &\begin{pmatrix} u_x \\ v_x
\end{pmatrix}=0, \lb{ex-stat} \\
n=1: \quad &\begin{pmatrix}   
(u+\beta v_x)_{xx} 
+4 \alpha (v(u+\beta v_x))_x-c_1 i(u+\beta v_x)_x \\
u_x+(\beta-\alpha) v_{xx} +2\alpha^2 (v^2)_x 
-2i c_1 \alpha v_x
\end{pmatrix}=0, \no \\
&\text{  etc.} \no
\end{align}
In the special homogeneous case, the latter set of 
equations, 
the stationary classical Boussinesq system, can be 
rewritten 
in the more familiar form
\begin{equation}
u+(\beta-\alpha)v_x+2\alpha^2 v^2=0, \quad 
(2\alpha\beta-\beta^2)v_{xxx}+(\alpha-\beta)u_{xx}
+4\alpha^2 (uv)_x=0. \lb{31}
\end{equation}
Using the first equation in \eqref{31}, the second can 
also be rewritten as $v_{xxx}-12\alpha^2(v^2)_x=0.$

To discuss the time-dependent hierarchy of classical 
Boussinesq
systems we follow the AKNS case and introduce a 
deformation parameter $t_n\in\bbC$ in the
functions $u$ and $v$, that is, $u=u(x,t_n)$, 
$v=v(x,t_n)$.
The time-dependent zero-curvature relation then reads
\begin{equation}
\U_{t_n}-\V_{n+1,x}+[\U,\V_{n+1}]=0,
\end{equation}
implying
\begin{align}
0&=\U_{t_n}-\V_{n+1,x}+[\U,\V_{n+1}] \lb{32a} \\
&=\left(\begin{smallmatrix} -\alpha
v_{t_n}+i\G_{n+1,x}+i\F_n-i(u+\beta v_x)\HH_n &
(u+\beta v_x)_{t_n}-i\F_{n,x}-2i(iz+\alpha v)\F_n+2i(u+\beta
v_x)\G_{n+1}
\\ i\HH_{n,x}-2i(iz+\alpha v)\HH_n+2i\G_{n+1} &
\alpha v_{t_n}-i\G_{n+1,x}-i\F_n+i(u+\beta v_x)\HH_n
\end{smallmatrix}\right). \no 
\end{align}
Using now the recursion \eqref{recurs} to compute 
$\ff_j$, $\g_j$, and
$\h_j$ for $j=0,\dots,n$, \eqref{32a} reduces to
\begin{align}
&\U_{t_n}-\V_{n+1,x}+[\U,\V_{n+1}] \lb{32b}\\
&\quad=\left(\begin{smallmatrix} -\alpha
v_{t_n}+i\g_{n+1,x}+i\ff_n-i(u+\beta v_x)\h_n & 
(u+\beta v_x)_{t_n} -i\ff_{n,x}-2i\alpha v \ff_n
+2i(u+\beta v_x)\g_{n+1} \\ 
i\h_{n,x}-2i\alpha v \h_n+2i\g_{n+1} & \alpha
v_{t_n}-i\g_{n+1,x}-i\ff_n+i(u+\beta v_x)\h_n 
\end{smallmatrix}\right)
=0, \no 
\end{align}
or equivalently,
\begin{subequations} \lb{33}
\begin{align}
\alpha v_{t_n}-i\g_{n+1,x}-i\ff_n+i(u+\beta v_x)\h_n&=0, 
\lb{33-a}\\
(u+\beta v_x)_{t_n}-i\ff_{n,x}-2i\alpha v \ff_n
+2i(u+\beta v_x)\g_{n+1} &=0, \lb{33-b} \\
\h_{n,x}-2\alpha v \h_n+2\g_{n+1}&=0. \lb{33-c}
\end{align}
\end{subequations}
Using $\h_{n,x}-2\alpha v \h_n+2\g_{n+1}=0$ in order 
to eliminate
$\g_{n+1}$ in \eqref{33-a} and \eqref{33-b},
then yields the following expressions for the 
time-dependent
classical Boussinesq hierarchy,
\begin{align}
&\alpha u_{t_n}-\f{i}{2}\beta \h_{n,xxx}+i\alpha\beta
v\h_{n,xx}-i\big((\beta+\alpha)u+\beta(\beta
-\alpha)v_x\big)\h_{n,x} 
\no\\  
&-i\big(\beta(u+(\beta-\alpha)v_x)_x-2\alpha^2 (u+\beta
v_x)v\big)\h_n+i(\beta-\alpha)\ff_{n,x}
-2i\alpha^2 v\ff_n=0, 
\lb{34} \\
&\alpha v_{t_n}
+\f{i}{2}\h_{n,xx}-i\alpha v\h_{n,x}
+i(u+(\beta-\alpha)v_x)\h_n-i\ff_n=0, \quad n\in\bbN_0. 
\no
\end{align}
For brevity, equations \eqref{34} will be denoted by
\begin{equation}
\cBsq_n(u,v)=0, \quad n\in\bbN_0. \lb{35}
\end{equation}

\begin{remark}
Observe that $\g_{n+1}$ defined by \eqref{33-c} does 
not satisfy
\eqref{recurs-c}, but rather
\eqref{33-a}.  This is in contrast to the stationary 
case as well as the
corresponding definitions for the AKNS hierarchy.  
As in the stationary
case, the
$n$th cBsq system contains $n$ integration constants 
$c_1,\dots,c_n$
when
$n\in\bbN$.
\end{remark}

Explicitly, the first few equations read
\begin{align}
\cBsq_0(u,v)&=\begin{pmatrix}\alpha u_{t_0}
-\alpha
 u_x \\
\alpha v_{t_0}-\alpha v_x \end{pmatrix}=0, \lb{ex1} \\
\cBsq_1(u,v)&=\left(\begin{smallmatrix} 
\alpha u_{t_1}-\f{i}{2}\alpha\beta v_{xxx}
+\f{i}{2}(\beta-\alpha)(u+\beta v_x)_{xx}
-2i\alpha^2 (uv)_x + c_1 (-\alpha u_x)\\
\alpha v_{t_1}-\f{i}{2}(u+\beta v_x)_x -2i \alpha^2 vv_x
+c_1 (-\alpha v_x)\end{smallmatrix}\right)=0, \no \\
& \text{  etc.} \no
\end{align}
In the homogeneous case $\cBsq_1(u,v)=0$ can be 
rewritten as
\begin{align}
\alpha u_{t_1}&=\f{i}{2}\beta(2\alpha-\beta)v_{xxx}
+\f{i}2(\alpha-\beta)u_{xx}+2i\alpha^2 (uv)_x, \no \\
\alpha v_{t_1}&=\f{1}2(\beta-\alpha)
v_{xx}+2i \alpha^2v v_x+\f{i}2u_x.
\end{align}
Finally, specializing to $\alpha=\beta$ one obtains the 
classical
Boussinesq system
\begin{equation}
u_{t_1}=\f{i}{2}\alpha v_{xxx}+2i\alpha (uv)_x, \quad
\alpha v_{t_1}=2i\alpha^2 v v_x+\f{i}{2}u_x.
\end{equation}
 
In the next section we provide a new proof of the fact 
that the
AKNS and the cBsq hierarchies are gauge equivalent
by exibiting an explicit gauge transformation between them. 
In the
last section this will  be used to
derive algebro-geometric solutions of the cBsq hierarchy.

%%%%%%%%%%%%%%%%%%%%%%%%%%%%%%%%%%%%%%%%%%%%%%%%%%%%%%%%%%
%--------- section
\section{The gauge equivalence of the cBsq and AKNS 
hierarchies} \lb{sect-equiv}
%%%%%%%%%%%%%%%%%%%%%%%%%%%%%%%%%%%%%%%%%%%%%%%%%%%%%%%%

We start by briefly recalling the effect of gauge 
transformations
on zero-curvature equations. Starting with the 
time-dependent
equations
\begin{equation}
\Psi_x=U\Psi, \quad \Psi_t=V\Psi, \quad \Psi
=\begin{pmatrix} 
\psi_1 \\ \psi_2 \end{pmatrix}, \lb{4.1}
\end{equation}
whose compatibility relation $\Psi_{xt}=\Psi_{tx}$ 
yields the
zero-curvature equation
\begin{equation}
U_t-V_x+[U,V]=0, \lb{4.2}
\end{equation}
we  introduce the gauge transformation
\begin{equation}
\PPsi =S\Psi, \quad S \text{ invertible.} \lb{4.3}
\end{equation}
Then one derives,
\begin{equation}
\PPsi_x=\U\,\PPsi, \quad \PPsi_t=\V\,\PPsi, \lb{4.4}
\end{equation}
with
\begin{equation}
\U=S_xS^{-1}+SUS^{-1}, \quad \V=S_tS^{-1}+SVS^{-1} 
\lb{4.5}
\end{equation}
and hence
\begin{equation}
\U_t-\V_x+[\U,\V]=0. \lb{4.6}
\end{equation}
The corresponding stationary formalism starts from
\begin{equation}
\Psi_x=U\Psi, \quad iy\Psi=V\Psi, \quad y\in\bbC 
\lb{4.7}
\end{equation}
and 
\begin{equation}
-V_x+[U,V]=0. \lb{4.8}
\end{equation}
The gauge transformation \eqref{4.3} then effects
\begin{equation}
\PPsi_x=\U\,\PPsi, \quad iy\PPsi=\V \, \PPsi, \lb{4.9}
\end{equation}
with 
\begin{equation}
\U=S_xS^{-1}+SUS^{-1}, \quad \V=SVS^{-1} \lb{4.10}
\end{equation}
and hence
\begin{equation}
-\V_x+[\U,\V]=0. \lb{4.11}
\end{equation}
Introducing the particular choice
\begin{equation}
S=\begin{pmatrix} (-p)^{1/2} &0 \\ 0 & 1/(-p)^{1/2} 
\end{pmatrix}, \quad p \text{ meromorphic on } \bbC \lb{4.12}
\end{equation}
and applying it to \eqref{4.7}--\eqref{4.11} in the case 
of the
stationary AKNS hierarchy, identifying $(U,V)$ and 
$(U,V_{n+1})$,
then yields the following result.

%%%%%%%%%%%%%%%%%%%%%%%%%%%%%%%%%%%%%%%%%%%%%%%%%%%%%%%%
\begin{theorem} \lb{t4.1}
The stationary AKNS and cBsq hierarchies are gauge 
equivalent 
in
the sense that
\begin{equation}
\U=S_xS^{-1}+SUS^{-1}, \quad \V_{n+1}=SV_{n+1}S^{-1}, 
\lb{4.13}
\end{equation}
with $(U,V_{n+1})$ and $(\U,\V_{n+1})$ given by 
\eqref{umatrix},
\eqref{24} and \eqref{u-matrix}, \eqref{v-matrix}, 
respectively,
and
$S$ defined by \eqref{4.12} {\rm (}with $p=p(x)${\rm )}. 
In particular, the pair $(u,v)$ given by
\begin{equation}
u(x)=-p(x)q(x)+\f{\beta}{2\alpha}\left(\f{p_x(x)}
{p(x)}\right)_x, 
\quad v(x)=-\f{1}{2\alpha}\f{p_x(x)}{p(x)}, \lb{4.14}
\end{equation}
satisfies the $n$th stationary cBsq system if and only 
if the
pair $(p,q)$, given by
\begin{align}
p(x)&=\exp\left(-2\alpha \int^x dx'\,v(x')\right), \no \\
q(x)&=-(u(x)+\beta v_x(x))\exp\left(2\alpha \int^x dx'\, 
v(x') 
\right), \lb{4.15}
\end{align}
satisfies the $n$th stationary AKNS system with identical 
sets of
integration constants $c_j\in\bbC$, $j=1,\dots,n$ 
for $n\in\bbN$.
\end{theorem} 
%%%%%%%%%%%%%%%%%%%%%%%%%%%%%%%%%%%%%%%%%%%%%%%%%%%%%%%%%%
\begin{proof}
$\U=S_xS^{-1}+SUS^{-1}$ is easily seen to be equivalent to
\eqref{4.14}. Similarly, $\V_{n+1}=SV_{n+1}S^{-1}$ is 
equivalent 
to
\begin{equation}
\F_n=-pF_n, \quad \G_{n+1}=G_{n+1}, \quad 
\HH_n=-\f{1}{p}H_n. 
\lb{4.16}
\end{equation}
Next suppose that $(p,q)$ solves the $n$th stationary 
AKNS system, that
is, equations \eqref{FGH} hold.  Define $\F_n$, 
$\G_{n+1}$, and $\HH_n$
by \eqref{4.16} and $(u,v)$ by \eqref{4.14}. Then 
clearly
\eqref{bar-FGH} is satisfied, proving that $(u,v)$ 
satisfies the $n$th stationary cBsq system.

Conversely, starting with $(u,v)$ solving the $n$th
stationary cBsq system  \eqref{bar-FGH}, we can 
define $(p,q)$ and $F_n$, $G_{n+1}$,
and $H_n$ using \eqref{4.15} and \eqref{4.16}, 
respectively. One 
then easily verifies that \eqref{FGH} holds, and thus 
$(p,q)$ solves the
$n$th stationary AKNS system.
\end{proof}
%%%%%%%%%%%%%%%%%%%%%%%%%%%%%%%%%%%%%%%%%%%%%%%%%%%%%%%%%%%

We note that the ambiguity inherent to \eqref{4.15}, due 
to an 
arbitrary integration constant, corresponds to the scale
invariance of the AKNS hierarchy as discussed in 
\eqref{28e}--\eqref{28g}.

The time-dependent analog of Theorem~\ref{t4.1} then 
reads as
follows.

%%%%%%%%%%%%%%%%%%%%%%%%%%%%%%%%%%%%%%%%%%%%%%%%%%%%%%%%
\begin{theorem} \lb{t4.2}
The time-dependent AKNS and cBsq hierarchies are gauge 
equivalent
in the sense that 
\begin{equation}
\U=S_xS^{-1}+SUS^{-1}, \quad \V_{n+1}=S_{t_n}S^{-1}+ 
SV_{n+1}S^{-1}, \lb{4.18}
\end{equation}
with $(U,V_{n+1})$ and $(\U,\V_{n+1})$ given by 
\eqref{umatrix}, 
\eqref{24} and \eqref{u-matrix} and \eqref{v-matrix}, 
respectively,
and 
$S$ defined by \eqref{4.12} {\rm (}with 
$p=p(x,t_n)${\rm)}. In particular, 
the pair $(u,v)$ given by
\begin{equation}
u(x,t_n)=-p(x,t_n)q(x,t_n)+\f{\beta}
{2\alpha}\left(\f{p_x(x,t_n)}{p(x,t_n)}\right)_x, 
\quad v(x,t_n)=-\f{1}{2\alpha}\f{p_x(x,t_n)}{p(x,t_n)}, 
\lb{4.19}
\end{equation}
satisfies the $n$th cBsq system $\cBsq_n(u,v)=0$ if and 
only if 
the pair $(p,q)$ given by 
\begin{align}
p(x,t_n)&=\exp\left(-2\alpha \int^x dx'\,v(x',t_n)\right), 
\no \\ 
q(x,t_n)&=-(u(x,t_n)+\beta v_x(x,t_n))
\exp\left(2\alpha \int^x
dx'\, v(x',t_n) \right), \lb{4.20}
\end{align}
satisfies the $n$th AKNS system $\AKNS_n(p,q)=0$ with 
identical
sets of integration constants $c_j\in\bbC,$ 
$j=1,\dots,n$ for $n\in\bbN$.
\end{theorem}
%%%%%%%%%%%%%%%%%%%%%%%%%%%%%%%%%%%%%%%%%%%%%%%%%%%%%%%%%%%
\begin{proof}
$\U=S_xS^{-1}+SUS^{-1}$ is equivalent to \eqref{4.19} as 
noted 
in the proof of Theorem~\ref{t4.1}. By a direct calculation, 
$\V_{n+1}=S_{t_n}S^{-1}+SV_{n+1}S^{-1}$ is equivalent to 
\begin{equation}
\F_n=-pF_n, \quad \G_{n+1}=G_{n+1}+\f{i}{2}\f{p_{t_n}}{p}, 
\quad \HH_n=-\f{1}{p}H_n. \lb{4.21}
\end{equation} 
Next assume that $(p,q)$ solves the $n$th  AKNS system, 
that
is, equations \eqref{28a} hold.  Define $\F_n$, 
$\G_{n+1}$, and
$\HH_n$ by \eqref{4.21} and $(u,v)$ by \eqref{4.19}. 
Then clearly
\eqref{30a} is satisfied, proving that $(u,v)$ 
satisfies the
$n$th cBsq system.

Conversely, starting with $(u,v)$ solving the $n$th
 cBsq system \eqref{30a}, we can define $(p,q)$ and 
$F_n$, $G_{n+1}$,
and $H_n$ using \eqref{4.20} and \eqref{4.21}, 
respectively.  Again 
one verifies that \eqref{28a} holds, and thus $(p,q)$ 
solves the $n$th
 AKNS system.
\end{proof}
%%%%%%%%%%%%%%%%%%%%%%%%%%%%%%%%%%%%%%%%%%%%%%%%%%%%%%%%%%%

The equivalence of the cBsq and AKNS hierarchies, on the 
basis of
the transformation \eqref{4.19} has first been noted by 
Jaulent 
and Miodek \cite{JM77} and later by Matveev and Yavor
\cite{MatveevYavor:1979}. It has been further discussed 
and linked
to Hirota's bilinear formalism by Sachs \cite{Sachs:1988}. 
Our
method of proof of Theorems~\ref{t4.1} and \ref{t4.2}, 
based
on the polynomial recursion formalism developed in
Section~\ref{sect-cBsq}, to the best of our knowledge, is 
new. 

%%%%%%%%%%%%%%%%%%%%%%%%%%%%%%%%%%%%%%%%%%%%%%%%%%%%%%%%%
%--------- section
\section{Algebro-geometric solutions of the classical 
Boussinesq
hierarchy} \lb{sect-algeb}
%%%%%%%%%%%%%%%%%%%%%%%%%%%%%%%%%%%%%%%%%%%%%%%%%%%%%%%%%%%

Finally we derive the theta function representation of 
algebro-geometric cBsq solutions utilizing the gauge 
equivalence of the cBsq and AKNS hierarchies. 

Let $(u^{(0)}, v^{(0)})$ be a stationary solution of the
$n$th classical Boussinesq system, that is,
\begin{align}
 &\ff_{n,x}+2\alpha v \ff_n
+(u+\beta v_x)(\h_{n,x}-2\alpha v\h_n)=0, \no \\
&\ff_{n}-(u+\beta v_x)\h_n-\f12\h_{n,xx}
+\alpha(v\h_n)_x =0, \lb{5.1}
\end{align}
for a given set of integration constants 
$\{c_j\}_{j=1,\dots,n}\subset\bbC$.  Fix
$r\in\bbN_0$ and corresponding integration 
constants $\{\tilde c_j\}_{j=1,\dots,r}
\subset\bbC$. The aim in this section 
is to construct a solution $(u,v)$ of 
\begin{equation}
\cBsq_r(u,v)=0, \quad (u,v)|_{t_r=t_{0,r}}=(u^{(0)}, 
v^{(0)}).
\end{equation}
The function $\pphi$ and the Baker--Akhiezer function 
$\PPsi$ 
associated with the
classical Boussinesq hierarchy can be obtained as 
follows.

%%%%%%%%%%%%%%%%%%%%%%%%%%%%%%%%%%%%%%%%%%%%%%%%%%%%%%%%%
%---------- theorem
\begin{theorem}\lb{theorem5.1}
Consider $P=(z,y) \in \calK_n 
\backslash \{P_{\infty_{\pm}}\}$ and  
$(z,x,x_0,t_r,t_{0,r}) \in \bbC^5$.
Let $\phi$, $\Psi$, and $S$ be given by \eqref{415},
\eqref{421}, and \eqref{4.12}, respectively. Define
\begin{equation}
\PPsi=\begin{pmatrix}\ppsi_1 \\ \ppsi_2\end{pmatrix}=S\Psi, 
\quad 
\pphi=-\f{\phi}{p}. \lb{BA}
\end{equation}
Then $\pphi(P,x,t_r)$  
satisfies  $\pphi=\ppsi_2/\ppsi_1$ and
\begin{multline}
v(x,t_r)\pphi(P,x,t_r)-\f{1}{2\alpha}\pphi_{x}(P,x,t_r) \\
+\f{1}{2\alpha}(u(x,t_r)+\beta v_x(x,t_r))\pphi(P,x,t_r)^2 
- \f{iz}{\alpha}\pphi(P,x,t_r)-\f{1}{2\alpha}=0. 
\label{424aa}
\end{multline}
$\PPsi(\dott,x,x_0,t_0,t_{0,r})$ is meromorphic on 
$\calK_n\backslash\{P_{\infty_\pm} \}$ and satisfies 
\begin{align}
\PPsi_x(P,x,x_0, t_r,t_{0,r})&=
\U(z,x,t_r)\PPsi(P,x,x_0,t_r,t_{0,r}),  \label{431A}\\
iy(P)\PPsi(P,x,x_0, t_r,t_{0,r})
&=\V_{n+1}(z,x,t_r)\PPsi(P,x,x_0, t_r,t_{0,r}), 
\lb{431Aaa} \\
\PPsi_{t_r}(P,x,x_0, t_r,t_{0,r})
&=\widetilde \V_{r+1}(z,x,t_r)\PPsi(P,x,x_0, t_r,t_{0,r}). 
\lb{431Aaaa}
\end{align}
\end{theorem}
%%%%%%%%%%%%%%%%%%%%%%%%%%%%%%%%%%%%%%%%%%%%%%%%%%%%%%%%%%%%%
%----------- proof
\begin{proof}  Immediate from Theorem \ref{l41} and 
\eqref{4.1}--\eqref{4.6}.
\end{proof}
%--------- end of proof 
%%%%%%%%%%%%%%%%%%%%%%%%%%%%%%%%%%%%%%%%%%%%%%%%%%%%%%%%%%%%

The explict representation of algebro-geometric solutions 
of the
classical Boussinesq hierarchy in terms of the Riemann theta
function associated with $\calK_n$ then reads as follows 
(we use 
the notation employed in Theorem~\ref{t2.2}).
%%%%%%%%%%%%%%%%%%%%%%%%%%%%%%%%%%%%%%%%%%%%%%%%%%%%%%%%%%%
%--------- theorem
\begin{theorem}\lb{theorem5.2}
Let $P\in\calK_n\backslash\{P_{\infty_\pm}\}$, 
$(x,x_0,t_r,t_{0,r})\in\bbC^4$,
assume $\calK_n$ to be nonsingular, and
suppose $\calD_{\ul {\hat \mu}(x,t_r)}$, or 
equivalently,
$\calD_{\ul {\hat \nu}(x,t_r)}$
to be nonspecial, that is,  
their index of speciality vanishes, 
$i(\calD_{\ul {\hat \mu}(x,t_r)})=
i(\calD_{\ul {\hat \nu}(x,t_r)})=0$. 
Moreover, suppose that $(u,v)$ satisfies $\cBsq_r(u,v)=0$  
with $(u,v)|_{t_r=t_{0,r}}=(u^{(0)},v^{(0)})$, a solution
of the $n$th stationary classical Boussinesq system.  Then 
the theta function representation of $(u,v)$ is given by
\begin{align}
u(x,t_r)&=e_{0,1}+
\f{\partial^2}{\partial x^2}\ln(\theta(\uz_
+(\hat {\ul \mu}(x,t_r)))) \no \\
&+\f{\beta}{2\alpha}
\f{\partial^2}{\partial x^2}\ln\left(
\f{\theta(\uz_-(\ul {\hat \nu}(x_0,t_{0,r})))}
{\theta(\uz_+(\ul {\hat \nu}(x_0,t_{0,r})))}
\f{\theta(\uz_+(\ul {\hat \nu}(x,t_r)))}
{\theta(\uz_-(\ul {\hat \nu}(x,t_r)))}\right),   \no \\
v(x,t_r)&=\f{i}{\alpha}e_{0,0}-\f{1}{2\alpha}
\f{\partial}{\partial x}\ln\left(
\f{\theta(\uz_-(\ul {\hat \nu}(x_0,t_{0,r})))}
{\theta(\uz_+(\ul {\hat \nu}(x_0,t_{0,r})))}
\f{\theta(\uz_+(\ul {\hat \nu}(x,t_r)))}
{\theta(\uz_-(\ul {\hat \nu}(x,t_r)))}\right).
\end{align}
\end{theorem}
%%%%%%%%%%%%%%%%%%%%%%%%%%%%%%%%%%%%%%%%%%%%%%%%%%%%%%%%%%%%
%-------- proof 
\begin{proof}
Combine Theorem~\ref{t2.2} and \eqref{4.19}. 
\end{proof}
%------- end of proof
%%%%%%%%%%%%%%%%%%%%%%%%%%%%%%%%%%%%%%%%%%%%%%%%%%%%%%%%%%%%

Obviously one can derive formulas similar to 
\eqref{465}--\eqref{467} for the
functions $\pphi$ and $\PPsi$ using the explicit relation 
\eqref{BA}. We leave the corresponding details to the reader.

Algebro-geometric solutions of the time-dependent classical 
Boussinesq system $\cBsq_1(u,v)=0$ and their theta function
representations were originally derived by Matveev and Yavor 
\cite{MatveevYavor:1979}. The case of real-valued solutions 
and 
additional reductions to elliptic solutions in the case of 
genus $n\leq 3$ were subsequently studied by Smirnov 
\cite{Smirnov:1986}. Theta function representations of
algebro-geometric solutions of $\cBsq_r(u,v)=0$ in the case 
$r\leq 3$ appeared in a recent preprint by Geng and Wu 
\cite{GengWu:1998}.

%%%%%%%%%%%%%%%%%%%%%%%%%%%%%%%%%%%%%%%%%%%%%%%%%%%%%%%%%%%%

%--------- acknowledgement
\bigskip

%%%%%%%%%%%%%%%%%%%%%%%%%%%%%%%%%%%%%%%%%%%%%%%%%%%%%%%%%%%
%{\bf Acknowledgments.}
%%%%%%%%%%%%%%%%%%%%%%%%%%%%%%%%%%%%%%%%%%%%%%%%%%%%%%%%%%%

%%%%%%%%%%%%%%%%%%%%%%%%%%%%%%%%%%%%%%%%%%%%%%%%%%%%%%%%%%%
%\bibliographystyle{PLAIN}
%\bibliography{solitonref}

\end{document}